\documentclass[fleqn,manuscript=article]{achemso}
\setkeys{acs}{usetitle=true} \usepackage[utf8]{inputenc} \usepackage{lineno}
\usepackage{bm} \usepackage{rotating} \usepackage[T1]{fontenc}
\usepackage{amsmath} \usepackage{amssymb} \usepackage{graphicx}
\usepackage[usenames,dvipsnames]{color}
\usepackage{comment,soul}

\usepackage[normalem]{ulem}

\usepackage{float} 

\title{Resist the surface field: the H-bond network decides if water aligns at metal electrodes}

\author{Mohammed Bin Jassar}
\affiliation{Laboratoire CPCV, Département de Chimie, École Normale Supérieure, PSL University, Sorbonne University, CNRS, 75005 Paris, France}

\author{Wei-tao Liu}
\affiliation{Physics Department, State Key Laboratory of Surface Physics, Key Laboratory of Micro and Nano Photonic Structures [Ministry of Education (MOE)], Fudan University, Shanghai 200433, China}
\email{wtliu@fudan.edu.cn}

\author{Simone Pezzotti}
\affiliation{Laboratoire CPCV, Département de Chimie, École Normale Supérieure, PSL University, Sorbonne University, CNRS, 75005 Paris, France}
\email{simone.pezzotti@ens.psl.eu}

\date{\today}
\begin{document}

\clearpage

\begin{abstract}
At an electrode, water molecules align to the surface field upon voltage application. This 
initiates important electrochemical reactions, e.g., hydrogen and oxygen evolution reactions. Recently developed non-linear optical techniques challenge the traditional picture by quantifying a lack of water alignment at metal electrodes. We here provide theoretical and experimental evidences for the existence of a driving force that opposes water alignment to the surface field. Such driving force originates from the ordering templated by the metal surface on the physisorbed water layer, and scales with surface hydrophilicity. We hence propose a physical model for water alignment at electrodes based on a balance of H-bond network and surface field driving forces, which reconciles the traditional picture with the new experimental observations. 
\end{abstract}

\clearpage
\begin{center}
{\bf \Large Graphical TOC Entry}\\[20mm]
\includegraphics[width=0.4\textwidth]{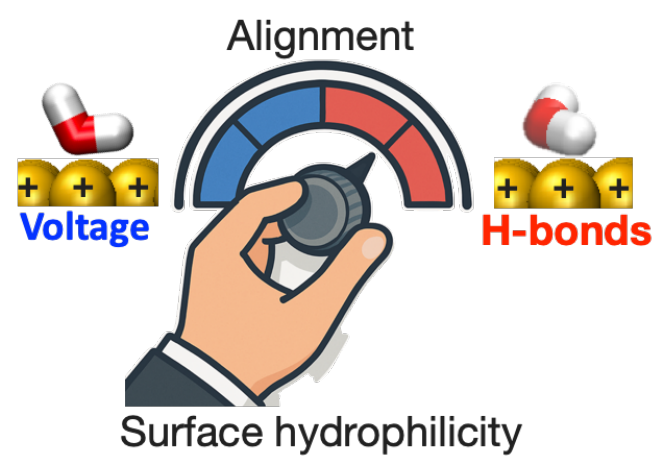}
\end{center}

\clearpage

\section{Introduction}\label{sec1}

The alignment of water molecules by the electric field at the metal electrode and the related restructuring of the H-bond network have been recognized to deeply influence the properties of the electric double layer, as well as the outcome of important electrochemical reactions, such as the hydrogen evolution reaction (HER) at the negative electrode and the oxygen evolution reaction (OER) at the positive electrode\cite{speelman2025quantifying, Gonella.2021, xu2023optical, li2022hydrogen, li2019situ, alfarano2021stripping, Staffell.2019, Zhao.2023, Murke.JACS2024, zhang2024situ, oer1,oer2,oer3}. There, properly aligning water at the electrode is the first step to initiate the chemistry.\cite{speelman2025quantifying, speelman2025water, fernandez2024effect,zhao2023action,liu2019unifying,xu2024cation,lam2020theory,monteiro2021understanding} 
For example, many recent studies have been dedicated to rationalizing the effect of spectator cations on HER measured electrochemically.\cite{li2022hydrogen, zhang2024situ,fernandez2024effect,zhao2023action,liu2019unifying,xu2024cation,lam2020theory,monteiro2021understanding} Several possible mechanisms have been proposed by combining experiments with molecular dynamics (MD) simulations. Despite their diversity, they share the belief that enhancement of HER depends on how ions restructure the water network at the negative electrode, changing the way water molecules are aligned, the H-bond connectivity between them, and the associated reorganization free energy.\cite{li2022hydrogen, zhang2024situ,fernandez2024effect,zhao2023action,xu2024cation,monteiro2021understanding} Similarly, in OER, water molecules first have to align with their oxygen atoms pointing towards the positive electrode to initiate the reaction; which was proposed to impact the efficiency.\cite{speelman2025quantifying, speelman2025water,liu2019unifying}. 

Characterizing such field-induced responses of water at the molecular level has become a central challenge in modern electrochemistry, as opposed to the historical mean-field view, in which water is reduced to a (local) dielectric. Yet, measuring the water orientation and H-bond network properties at the electrode with molecular resolution remains a hard nut to crack.

 The interfacial water network can be probed in operando by vibrational spectroscopies, such as Surface-Enhanced Raman (SERS)\cite{li2019situ, Shi2018} and IR (SEIRAS)\cite{Zhao.2023, Osawa_JACS2013} , and difference Terahertz (THz) spectroscopy\cite{alfarano2021stripping}. These methods probe different aspects of the water network: SERS and SEIRAS focus mostly on OH stretching vibrations, while THz probes intermolecular hydrogen-bond stretching and libration modes. For instance, THz spectroscopy combined with MD simulations identified ordered water adlayer structures templated by strong interacting metal surfaces, where water molecules maximize intra-adlayer H-bonds within themselves.\cite{alfarano2021stripping, limmerpnas, pnasgold} The intra-adlayer H-bond stretching modes are red-shifted compared to the 200 $cm^{-1}$ H-bond stretching band in bulk water, allowing their spectroscopic identification. As a trade-off, these techniques lack direct information about water dipole orientation.

Coherent nonlinear optical spectroscopies, including Second Harmonic Generation (SHG) and Sum Freuqency Generation (SFG), can be surface-specific and decisive in overcoming this challenge,\cite{cowan2019vibrational, nihonyanagi2004potential, piontek2023probing, w1, w2} in particular when combined with theoretical spectroscopy from MD simulations. Recently, SHG was pioneered by Geiger et al.\cite{speelman2025quantifying} for counting the number of water molecules aligned by the applied voltage at a Ni electrode. By leveraging the direct relationship between the measured Im$\chi^{(2)}$ response and the net water dipole orientation, the study quantified the progressive alignment of water upon surface charging and estimated the associated flipping energetics. Surprisingly, they reported a lack of water alignment in a $\sim$0.8 V range. The underlying changes in the interfacial water H-bond network could not be identified since this information in not provided by SHG. Using surface plasmon enhanced sum frequency vibrational spectroscopy, Liu et al.\cite{w1} probed the voltage induced changes in the interfacial water spectrum near a gold electrode. Again, there appeared a potential range indicating the lack of water alignment, which calls for further insight into the molecular-level structure at the electrified interfaces.

Here, we adopt an integrated approach that combines plasmon-enhanced SFG spectroscopy experiments\cite{w1,w2} with a recently introduced pop-model method for theoretical SFG spectroscopy~\cite{PopModel_2024} and structural characterization from constant potential classical MD simulations, at an electrified gold/water interface. SFG measures the second order susceptibility $\chi^{(2)}(\omega)$, which is zero in the centrosymmetric bulk and non-zero at the interface, where centrosimmetry is broken. At a fixed number of OH bonds, the amplitude of $\chi^{(2)}(\omega)$ in the OH-stretching frequency range deoends on the water dipole orientation. The more water OH groups point along the surface normal, the larger the $\chi^{(2)}(\omega)$ contributed by their stretching modes. The spectral shape of $\chi^{(2)}(\omega)$ informs about the H-bonds formed by interfacial water. The pop-model method for theoretical SFG spectroscopy allows us to quantitatively relate the spectroscopic and structural properties of the  interface. With that, we uncover new molecular insights into the physics of the interfacial water response to the electric field: the field-alignment of interfacial water molecules upon both positive and negative surface charging is hampered by the ordering templated by the metal surface into the interfacial water network. We rationalize this in terms of a free energy cost to restructure the H-bond network within the physisorbed water adlayer, which scales with surface hydrophilicity. This H-bond network driving force may shed new light on the origin of water free energy contributions to both HER and OER.

\section{Results and Discussions}\label{sec2}

 \textbf{How does water align to the applied voltage? 
}

\begin{figure}[H]
\begin{center}
\includegraphics[width=0.95\textwidth]{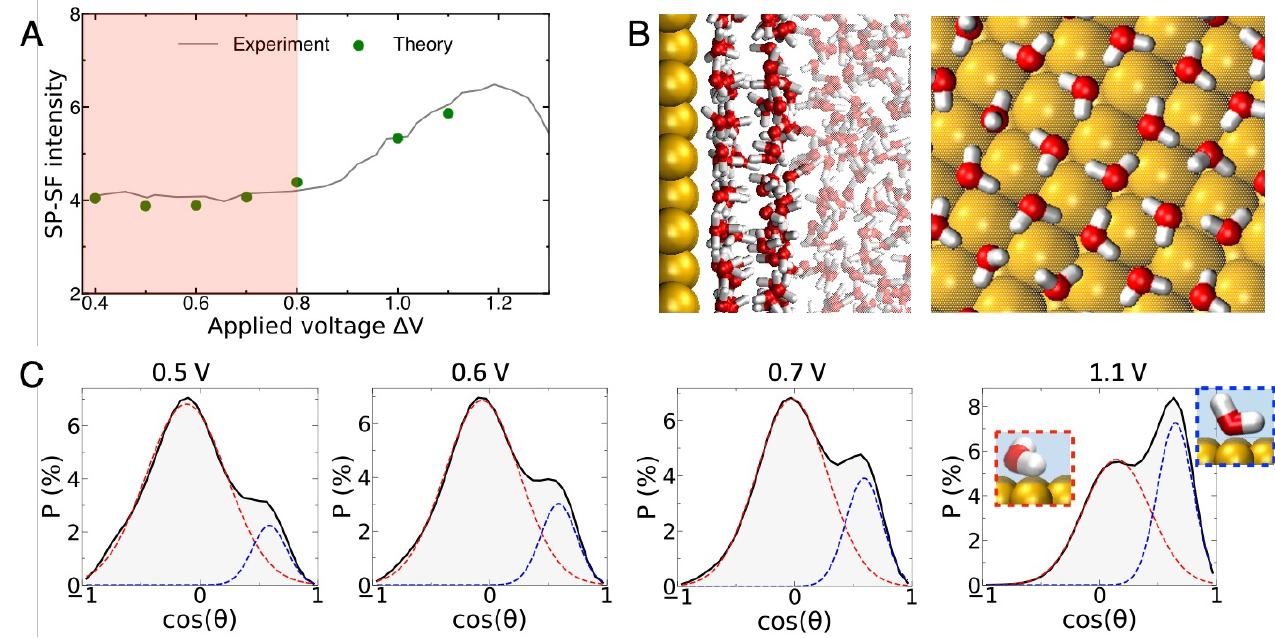}
\end{center}
\caption{Water does not align to the surface field at a Au electrode. (A) Comparison between measured (solid curve) and predicted (green dots) PS-SFG intensity (in the OH-stretching range) as a function of applied voltage. The red area identifies a $\sim$0.5 V wide region around the PZC (0.6 V) where the intensity - which is sensitive to the water dipole orientation - does not change with V.  (B) MD-snapshot illustrating the water structure at the interface responsible for the lack of water orientation. The top-view shows that adlayer water molecules preferentially form intra-layer H-bonds between themselves, oriented flat on the surface. (C) Distribution of water dipole orientation ($\cos \theta$) for different applied voltages ($\Delta \text{V}$), from which the SFG intensity is computed according to Eq. \ref{eq:SFG-ADL-diff}.}\label{effects}
\end{figure}

Fig. \ref{effects}A displays the measured SFG intensity (in the OH-stretching range) as a function of applied voltage at a gold/water interface. Strikingly, we observe that within a ±0.25 V window around the potential of zero charge (PZC = 0.6 V), the SFG intensity remains minimal and nearly constant, see red region of Fig. \ref{effects}A. This behavior is inconsistent with the conventional description  of water response to an electric field, as if water is ignoring the surface electric field and does not align to it. Remarkably, it holds true both above and below the PZC. This result is similar to a recent SHG study from Geiger et al.\cite{speelman2025quantifying} at a Ni/water interface, where water molecules were found not aligned at low voltages around the open circuit potential (OCP). 
When further increasing the applied voltage beyond 0.9 V, we observe a gradual rise in SFG intensity, consistent with the textbook water response aligning to the electric field. A further voltage increase, above 1.2 V, leads to a drop in SFG intensity, which was previously attributed to gold surface oxidation.\cite{w1,w2} We note that lower voltages that 0.3 V could not be measured because of bubbles formation at the interface due to the onset of interfacial chemistry, as detailed previously.\cite{w1,w2} 
Herein, our analysis focuses on the 0.4-1.2 V region, where changes in SFG intensity is dominated by the field-driven water reorientation, with little effect from surface chemistry.

To better understand the lack of water alignment in the experiments, we turn our attention to the voltage-induced structural changes at the interface.  We characterize these changes from constant-potential MD simulations (see methods for computational details), and we quantitatively relate the orientation of water molecules as a function of $\Delta$V to the measured SP-SFG intensity by means of the pop-model for theoretical SFG spectroscopy.\cite{PopModel_2024} Since in our experiment we used plasmonic nanogratings to enhance the SFG response, we expect that the recorded intensity is dominated by the contribution from the $N_{ADL}$ adlayer water molecules in direct contact with the Au surface (illustrated in Fig. \ref{effects}B). We hence focus our theoretical analysis on these water molecules. The same trends as discussed hereafter are also observed by increasing the thickness over which water contributions are considered (see Fig.S1 of SI). 

To understand the molecular origin of the lack of SFG intensity changes at 0.4-0.9 V (red region), we start by calculating the theoretical SFG response due to water dipoles reorientations within the adlayer using a variant of the recently introduced pop-model approach.\cite{PopModel_2024}. In a nutshell, the SFG response from the adlayer water is expressed by Eq.  \ref{adla}:

\begin{equation}\label{adla}
\chi^{(2)}_{ADL}(\omega)= \frac{i\omega}
      {k_BT} \sum_{i=1}^{N{_{ADL}}} \left( \sum_{j=1}^{N{_W}} \int_0^{\infty}dt e^{i\omega t}\langle \alpha_{xx}^j(t) \mu_{z}^i(0)\rangle \right)  = \sum_{i=1}^{N{_{ADL}}} \beta_{i}(\omega)
\end{equation}
where N$_{ADL}$ refers to the water molecules located within the adlayer only (at each time 0). The $\frac{i\omega}
      {k_BT}\sum_{j=1}^{N{_ADL}} \int_0^{\infty}dt\exp(i\omega t)\langle \alpha_{xx}^j(t) \mu_{z}^i(0)\rangle )$ term in the equation represents the contribution to the total SFG spectrum of the i-th water molecule in the adlayer, labeled $\beta_{i}(\omega)$, which includes all the cross-terms with surrounding N$_W$ water molecules. $\beta_{i}(\omega)$ has the dimension (and macroscopic definition) of a molecular hyperpolarizability. Hereafter, $\chi^{(2)}_{ADL}$ and $\beta_{i}$ are written by omitting ($\omega$) for simplicity. SFG intensity variations upon surface charging are generally dominated by the reorientation of water molecules.\cite{wen2016unveiling, Roke_JPCC_2016, Morita_PCCP_2018, Tahara_JPCL_2018, Pezzotti_PCCP2018, PopModel_2024, borghetto2023oxide, gibbs2022water, gonella2021water, speelman2025water} This is explicated in Eq.  \ref{eq:betaADL} by multiplying and dividing $\beta_i$ by the orientation of each i-th water molecule, $cos\theta_i$, following the derivation of ref.\citenum{PopModel_2024}:

\begin{equation}\label{eq:betaADL}
\chi^{(2)}_{ADL} = \sum_{i=1}^{N{_{ADL}}} \frac{\beta_{i}}{cos\theta_i}~ cos\theta_i = \sum_{i=1}^{N{_{ADL}}} \beta_i^{eff}~ cos\theta_i
\end{equation}

where $\theta_i$ is the angle between the water dipole direction and the surface normal. 
This formulation captures the reorientation of interfacial water molecules that primarily drives the observed SFG intensity response\cite{wen2016unveiling, Morita_PCCP_2018, Pezzotti_PCCP2018, Tahara_JPCL_2018, borghetto2023oxide, gibbs2022water, gonella2021water}.  The same equation also describes the SHG signal (with a different $\beta_i^{eff}$).\cite{speelman2025quantifying}

    In the pop-model approach, Eq.\ref{eq:betaADL} is used to compute the theoretical SFG spectrum by identifying interfacial water populations (based on the local coordination of the water molecules). Each population has a different $ \beta_i^{eff}$, which was parametrized in a database from several aqueous interfaces.\cite{PopModel_2024, Wanlin_PRR2025} The total SFG spectrum is hence computed by  calculating the $\sum_{i=1}^{N_{ADL}} \cos\theta_i$ term for each identified population from the MD simulations of the system of interest (see ref.\citenum{PopModel_2024} for all details). Building on many previous studies \cite{wen2016unveiling, Pezzotti_PCCP2018, Morita_PCCP_2018, Tahara_JPCL_2018, borghetto2023oxide, gibbs2022water, gonella2021water, PopModel_2024, Roke_JPCC_2016, speelman2025water}, we anticipate that the voltage-dependent SFG intensity change is dominated by the water orientation term ($\sum_{i=1}^{N_{ADL}} \cos\theta_i$) and  has a comparably negligible dependence on $\beta_i^{eff}$. Therefore, to start, we simplify Eq.\ref{eq:betaADL} by just considering a single, average $\beta_i^{eff}$ for all adlayer water molecules, and by assuming $\beta_i^{eff}(V) = \beta^{eff}(PZC)$ . This leads to:
\begin{equation}\label{eq:SFG-ADL-diff}
\chi^{(2)}_{ADL}(V) - \chi^{(2)}_{ADL}(PZC) = \beta^{eff}(PZC) (\sum_{i=1}^{N{_{ADL}}} cos\theta_i(V) - \sum_{j=1}^{N{_{ADL}}}  cos\theta_j(PZC))
\end{equation}
where $\sum_{j=1}^{N_{ADL}} \cos\theta_j(PZC)$ is obtained from our classical MD simulations, while we take $\chi^{(2)}_{ADL}(PZC)$ as equal to the experimentally measured SP-SFG intensity at PZC. $\beta^{eff}(PZC)$ acts as a scaling factor that we fit to the experimental data. This approach inherently includes experimental factors like Fresnel coefficients and surface contributions in the scaling parameters $\chi^{(2)}_{ADL}(PZC)$ and $\beta^{eff}(PZC)$. The good agreement between theoretical and experimental results, shown in Fig. \ref{effects}A, provides an a posteriori validation that the voltage-dependent changes in SP-SFG intensity are indeed dominated by the reorientation of adlayer water. Most important, it allows us to quantitatively interpret the lack of spectroscopic changes in the 0.4--0.9 V range in terms of the field-induced water alignment as characterized from the MD simulations (Fig. \ref{effects}C, where $\cos \theta$ defines the water dipole orientation along the surface normal).

At PZC, the  $\cos \theta$ distribution is bimodal, with a preferred dipole orientation at $\cos \theta \simeq 0$ (i.e. parallel to the surface, red dashed curve) and a second preferred orientation at  $\cos \theta \simeq 0.5$, i.e. pointing away from the surface (blue). Strikingly, the $\cos \theta \simeq 0$ population remains dominant and nearly unvaried within a ±0.2 V window around the PZC. Analysis of the MD simulations further reveal that these water molecules are most commonly oriented flat on the surface ( as shown by the snapshot in the dashed-red box of Fig. \ref{effects}C, as well as in the top-view of Fig. \ref{effects}B), with both OH-groups forming intra-adlayer H-bonds. Only at higher voltages, at 1.1 V, water molecules start to lose the horizontal ordering (decrease of the probability maximum at $\cos \theta \simeq 0$) and align to the surface field (increased probability at $\cos \theta \simeq 0.5$), by reorienting on average one OH group pointing away from the surface (as illustrated in the snapshot in the dashed-blue box). We report in Figs.S2 and S3 of the SI additional theoretical SFG spectra for varying surface geometry (100 $vs$ 111) and ionic strength (no ions $vs$ 0.1 M NaCl) to show the generality of our findings.

\begin{figure}
    \centering
    \includegraphics[width=1\linewidth]{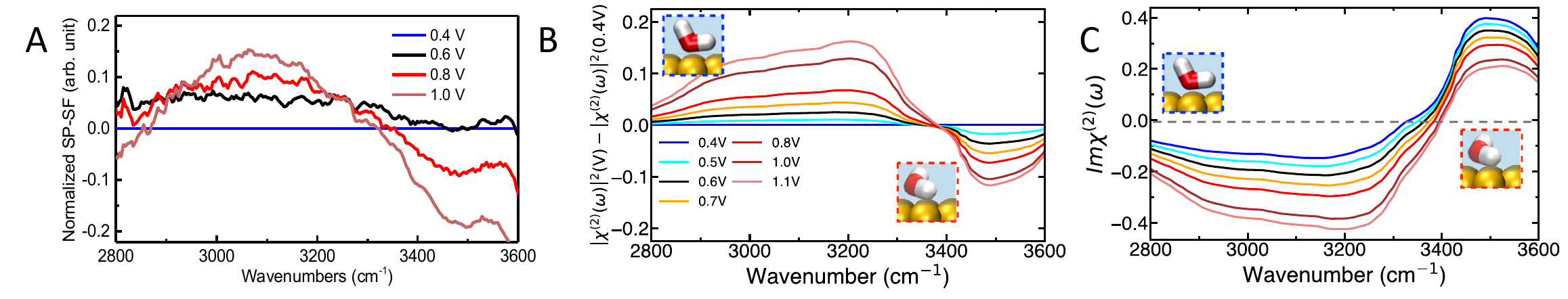}
    \caption{The culprit is the H-bond network. (A) Difference in the normalized (see text) SP-SFG spectra at various applied voltages during the anodic scan, with respect to 0.4 V. (B) Corresponding difference theoretical spectra. The schemes illustrate the interfacial water populations responsible for the two observed spectral features that change with applied voltage. (C) Theoretical $Im\chi^{(2)}(\omega)$ spectra, showing the same trends, with a clearer connection to water orientation (positive/negative $Im\chi^{(2)}(\omega)$ corresponds to water OH groups pointing toward surface/bulk water). As discussed in the text, these changes are signatures of the interfacial H-bond network restructuring required to align water to the surface field.}
    
    \label{sfg_spectra_shape}
\end{figure}

\textbf{The H-bond network is responsible for the lack of water alignment. } To gain a deeper understanding of the underlying changes occurring within the interfacial H-bond network, we focus now on the variations in spectral shape. The experimental spectra adapted from ref.\citenum{w1} are plotted in Fig.~\ref{sfg_spectra_shape}A. These were obtained by performing a double normalization process, with details provided in ref.\citenum{w1}. Briefly, we divided Au/H$_2$O spectra by the integrated intensity of Au/D$_2$O spectra at corresponding potentials, and then against the Au/water spectrum at 0.4 V vs. RHE, to remove spectral dependence from both IR input and local field factors. Variation in the normalized spectra then indicates the change in OH spectral profile. As the voltage increases, we observe a gain in intensity at low frequencies, around 2900-3200 cm$^{-1}$, and a loss of intensity at higher frequencies, $\sim$3400-3600 cm$^{-1}$. Similar spectral changes are typically interpreted as an increase in the ordering of the interfacial H-bond network, by associating the band at  $\sim$2900-3200 cm$^{-1}$ to more strongly H-bonded water molecules compared to the $\sim$3400-3600 cm$^{-1}$ band. 

To  interpret these spectral changes at a molecular level, we employ the pop-model approach as defined by Eq.  \ref{eq:betaADL} to compute the SFG spectrum of the interface as a weighted sum of the contributions from the two interfacial water populations identified from the MD simulations. As illustrated in Fig. \ref{effects}C, these correspond to the two features in the water $\cos \theta$ profiles (highlighted in blue and red). Population 1 (red) consists of water molecules that donate on average two intra-adlayer H-bonds (while accepting one H-bond from water in the adjacent layer). These molecules have low SFG activity due to their almost flat orientation: only a small component of their OH stretching modes that have the transient dipole pointing slightly toward the Au surface is detected by SFG. These modes are responsible for the $\sim$3400-3600 cm$^{-1}$ band in the theoretical spectra of Fig.~\ref{sfg_spectra_shape}B. This band is blue-shifted compared to the OH-stretching signature of bulk water, indicating apparently weaker H-bonds. This result nicely agrees with previous, complementary observations in the THz frequency range,\cite{Alfaranoe2108568118} where the H-bonds formed between adlayer water molecules were found to vibrate at lower frequency than the H-bonds formed in bulk water (please note that lower frequency for H-bond stretching modes in the THz range corresponds to higher frequency for OH stretching modes in the mid-IR range), which was ascribed to reduced cooperativity within the adlayer water network compared to the 3-dimensional network of bulk water. Population 2 (blue) contains water molecules that are preferentially oriented with one OH group involved in an intra-adlayer H-bond and the other OH-group donating a H-bond to water in the adjacent bulk-like water layer. This population is commonly observed at aqueous interfaces,\cite{PopModel_2024} and is responsible for the  $\sim$2900-3200 cm$^{-1}$ band in the theoretical spectra of Fig.~\ref{sfg_spectra_shape}B.

The theoretical voltage dependent SFG spectra of Fig.~\ref{sfg_spectra_shape}B are hence obtained as the weighed sum over these two partial contributions, where the voltage-dependent weighting factors $\sum_{j=1}^{N_{population}} \cos\theta_j(V)$ for the two populations are directly computed from the MD simulations. The theoretical spectra reproduce the experimental trends (only a qualitative comparison is made here due to the normalization method for experimental spectra detailed above). This allows a molecular-level interpretation of the changes within the interfacial water H-bond network responsible for the spectroscopic observable. As the voltage increases, a fraction of adlayer water molecules reorient from population 1 to population 2, causing the observed apparent red-shift in the experimental spectra. This red-shift, therefore, does not indicate an increase in water ordering but rather reflects a reconstruction of the interfacial water network, from a horizontal to a vertical ordering. This reconstruction occurs progressively with increasing voltage, until the fraction of aligned water molecules overcomes that of horizontally oriented ones.
The same trends are also observed in the theoretical $Im\chi^{(2)}(\omega)$ spectra of Fig.~\ref{sfg_spectra_shape}C, where it is easier to appreciate the distinct orientation of the two water populations (positive/negative $Im\chi^{(2)}(\omega)$ corresponds to water OH groups pointing toward surface/bulk water).

\begin{figure}[h]
    \centering
    \includegraphics[width=1\linewidth]{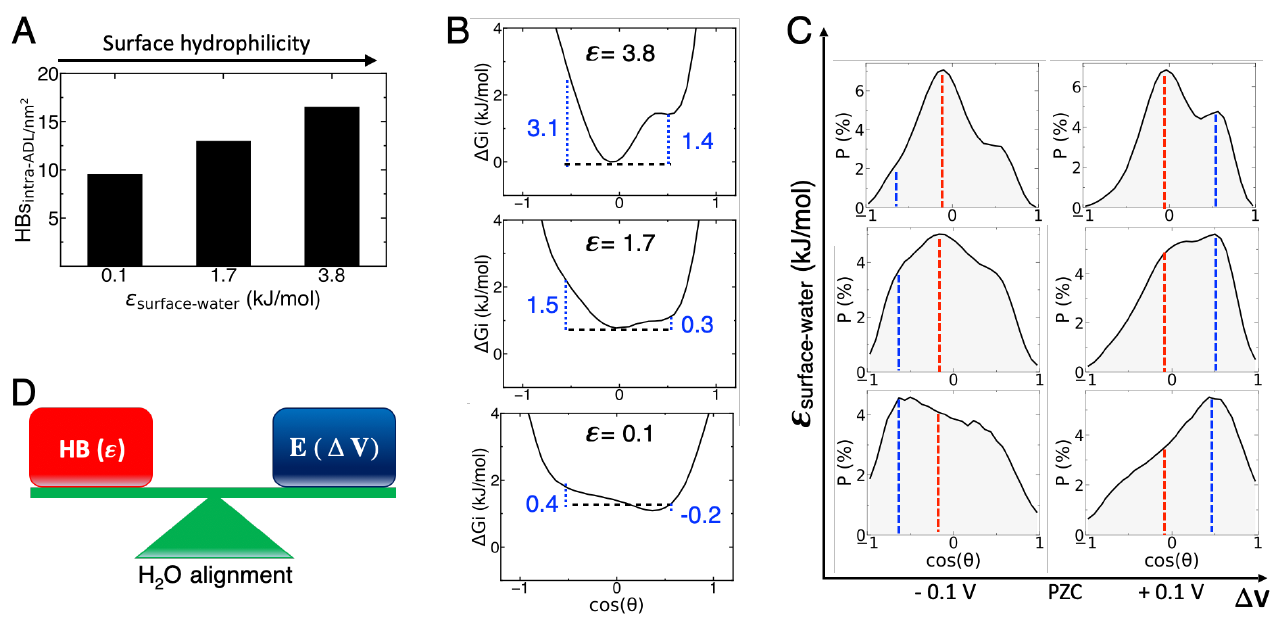}
    \caption{A theoretical experiment reveals a balance of driving forces. (A) The horizontal ordering templated by the surface on the interfacial HB-network, quantified by the density of intra-adlayer HBs between adlayer waters (top-view of Fig.\ref{effects}B), is reducedx by decreasing the surface-water interaction potential, $\epsilon_{surface-water}$. This corresponds to reducing surface hydrophilicity (SI, Table S1).\cite{BinJassar_amphiphilic2025} (B) Water dipole orientation, $\cos\theta$, free energy profiles at PZC, from which we quantify the free energy penalty $\Delta G_{HB}$ to restructure interfacial HB-network and align water upon negative/positive applied voltage, as illustrated. From top to bottom, $\Delta G_{HB}$ decreases with $\epsilon_{surface-water}$. (C) Corresponding changes in water alignment upon positive and negative applied voltages, for $\pm 0.1 V$ as an example. The dashed lines show that water is more easily aligned with decreasing $\epsilon_{surface-water}$, in agreement with Eq.\ref{eq:model}. (D) From the theoretical experiment, we conclude that water alignment is dictated by a subtle balance of driving forces from surface field (E, which scales with $\Delta V$) and surface induced horizontal ordering within the HB network (which scales with surface-water interactions).}
    \label{third}
\end{figure}

Based on these results, we hypothesize the existence of a H-bond network penalty ($\Delta G_{HB}$) that has to be overcome for the surface field to align interfacial water. Hence, the free energy of water orientation can be written as the sum of two opposing contributions:
\begin{equation}\label{eq:model}
\Delta G(\cos \theta) = \Delta G_{HB} (\cos \theta, \epsilon) + \Delta G_{field} (\cos \theta, \Delta V)
\end{equation}
where the $\Delta G_{HB}$ penalty (to align water dipoles to $cos \theta$) originates from the horizontal ordering induced by the strongly interacting metal surface on the water adlayer.\cite{pnasgold, limmerpnas} According to recent studies, the degree of horizontal ordering templated by the surface is a function of surface hydrophilicity, i.e. of surface--water interaction strength, $\epsilon$ (as well as of surface topology/morphology).\cite{gading2024role, jassar2024chemistry, BinJassar_amphiphilic2025, limmerpnas, rotenberg2011molecular} Hence, $\Delta G_{HB} (\epsilon)$ must be a function of surface hydrophilicity, too. Instead, the $\Delta G_{field}(\Delta V)$ driving force to align water to the surface field is a function of applied voltage. The physical model described by eq.\ref{eq:model} implies that water aligns to the surface field only above the threshold voltage $V^*$ that satisfies the condition $\Delta G_{field} = - \Delta G_{HB}$; $V^*$ being dependent on surface-water interactions, $\epsilon$. From fig.\ref{adla}, $V^* \simeq$ +0.25 V relative to PZC at the positive Au electrode.

In Fig.\ref{third}, we test this hypothesis with an ad-hoc designed theoretical experiment.  This experiment leverage on a recent study where we found that the degree of horizontal ordering that a surface induces within the adlayer (as well as its hydrophilicity) can be effectively tuned by adjusting the surface-water interaction strength (for a given topology) in the MD simulations.\cite{BinJassar_amphiphilic2025} Hence, we take our Au surface and reduce its interaction potential with water $\epsilon_{surface-water}$ (see methods), from the original value of 3.8 kJ/mol for Au, to 1.7 kJ/mol and to 0.1 kJ/mol. Fig.\ref{third}A indeed confirms that by reducing $\epsilon_{surface-water}$ we decrease the horizontal ordering within the adlayer, as quantified by the density of intra-adlayer water-water H-bonds formed between adlayer wayer molecules. We further computed theoretical contact angles (SI, Table S1) to verify the corresponding reduction in surface hydrophilicity. We then examine how the surface modification affects water alignment. Fig. \ref{third}B reports water dipole orientation free energy profiles as a function of $\cos\theta$ at PZC (computed from the $\cos\theta$ probability distributions according to $\Delta G (\cos \theta)= - ln (P(\cos \theta)/k_BT)$). We hence quantify the free energy penalty $\Delta G_{HB}$ to align water, as illustrated in the figure. In particular, we calculate the $\Delta G_{HB}$ penalty for aligning water to $\sim |cos\theta| =  0.5$, corresponding to the two interfacial water populations that emerge upon voltage application in our MD simulations, with one OH field-aligned only and the other still involved in an intra-adlayer H-bond. Full dipole alignment to $cos\theta = \pm 1$ costs much more in terms of free energy and is not statistically relevant in the explored voltage range. The $\Delta G_{field}$ values at different applied voltages (that oppose $\Delta G_{HB}$ according to eq.\ref{eq:model}) are quantified in Fig.S4, for comparison. In the applied voltage range where $\Delta G_{HB}>\Delta G_{field}$, we find the empirical linear relationship $\Delta G_{field} = C ~\Delta V \cos \theta$ (for both positive and negative surface charging). This allows us to derive an analytic expression for the threshold voltage $V^*$  (at which $\Delta G_{field} = - \Delta G_{HB}$) :
\begin{equation}\label{eq:V*}
V^* = \frac{\Delta G_{HB} (\cos \theta, \epsilon)}{C \cos \theta}
\end{equation}
with C = -11.7 ~kJ/mol/V.  In agreement with our hypothesis, Fig. \ref{third}B displays the largest $\Delta G_{HB}$ (hence $V^*$) values for the most hydrophilic surface. Moreover, we observe an asymmetric behavior, with a larger $\Delta G_{HB}$ for reorienting water toward negative $\cos \theta = - 0.5$ ($\Delta G_{HB} = 3.1 ~kJ/mol$, giving $V^* = -$ 0.52 V) then toward positive  $\cos \theta = + 0.5$ ($\Delta G_{HB} = 1.4 ~kJ/mol$, $V^* =+$ 0.23 V, in excellent agreement with the experiments), i.e., it is easier to reorient water at the positive then at the negative electrode. This asymmetry, which persists with decreasing interaction strength, is consistent with previous studies on carbon electrodes.\cite{laage2020graphene, olivieri2024water} Both $\Delta G_{HB}$ values vanish for the weakest interacting surface, where $V^*\simeq0~V$. This nicely explains why the lack of water alignment along the surface field has never been observed in the previous studies on carbon electrodes.\cite{laage2020graphene, wang2023direct, olivieri2024water} The surface modulation of the $\Delta G_{HB}$ driving force determines the interfacial water response upon electrode charging, as showcased in Fig. \ref{third}C for $\pm$0.1 V as an example. As evident in the cosine probability distributions, with decreasing  $\epsilon_{surface-water}$, interfacial water progressively aligns to both the positive and negative electrode, until the textbook behavior is recovered for the most weakly interacting surface (where  $V^* \simeq 0 ~V$). From our theoretical experiment, we hence conclude that water alignment at the electrode is not a one-dimensional, but a complex (at least) two-dimensional problem, where the final net alignment results from a subtle balance of surface field and surface induced ordering within the water H-bond network, as described by eq.\ref{eq:model}.

\begin{figure}[h]
    \centering
    \includegraphics[width=1\linewidth]{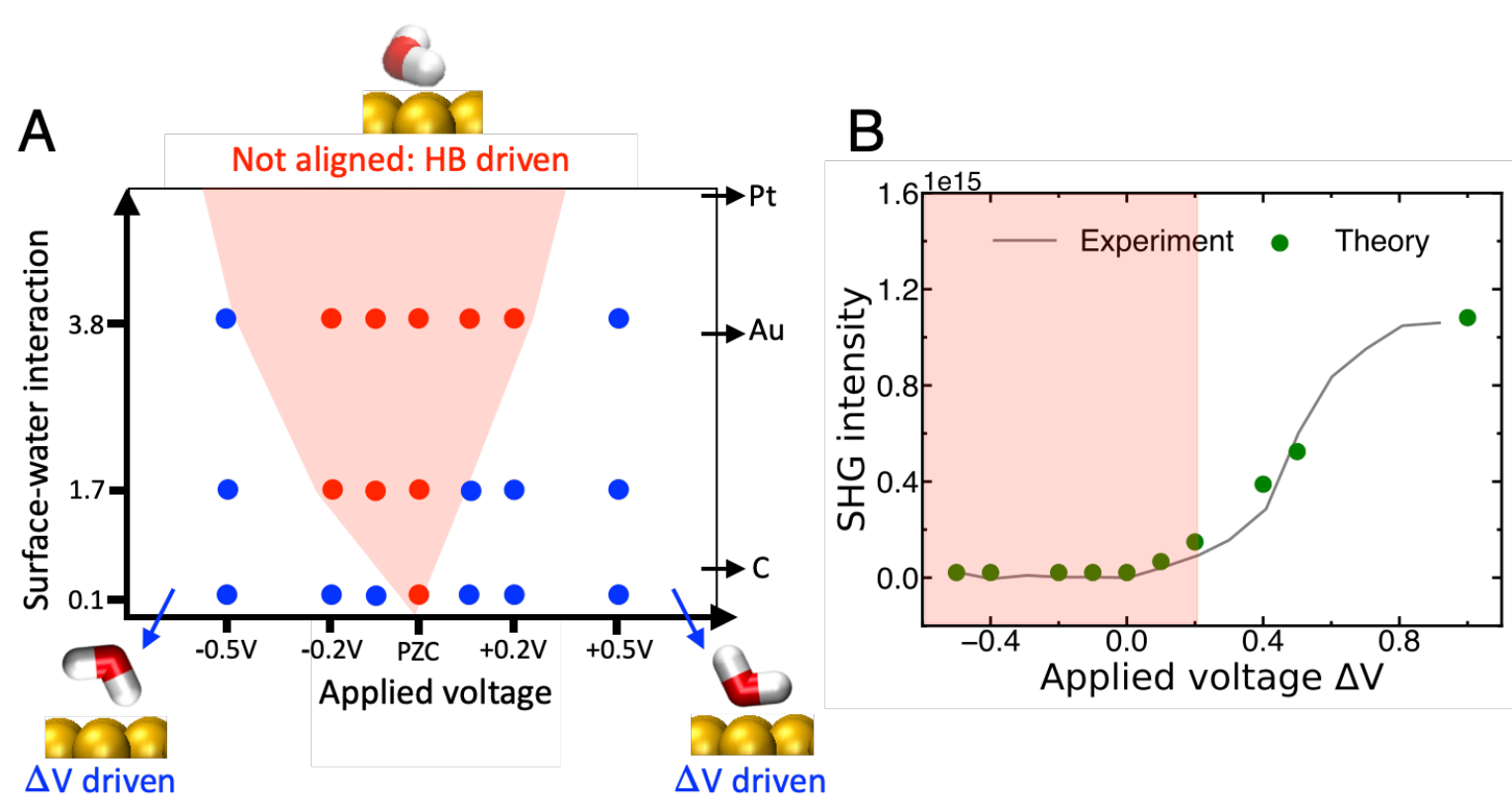}
    \caption{Beyond metal electrodes: a generalized physical model for water alignment. (A) Phase diagram of water alignment at electrified solid-liquid interfaces. The x,y-axis tune the identified driving forces from surface field ($\Delta G_{field}$) and surface induced ordering within the H-bond network ($\Delta G_{HB}$), respectively. The points are simulated conditions, colored according to the preferential water orientation: flat (red) or aligned (blue). The range of interaction strengths typical of Carbon, Au and Pt electrodes are marked. The red-shaded area illustrates the domain where $\Delta G_{HB}$ dominates and water does not align. It vanishes for low interacting surfaces, where water orientation simplifies to the textbook one-dimensional function of applied voltage. (B) With this understanding, we quantitatively model the recent SHG experiments from Geiger et al.\cite{speelman2025quantifying} at Ni/water via Eq. \ref{eq:conditioned}. Here, 0 V corresponds to OCP, not PZC.} 
    \label{surface_effect}
\end{figure}

\textbf{A generalized understanding on water alignment at solid-liquid interfaces.} In Fig.\ref{surface_effect}A, we formulate this understanding in a phase diagram for water alignment, where the surface and applied voltage dimensions tune the identified $\Delta G_{HB}(\epsilon)$ and $\Delta G_{field}(\Delta V)$ driving forces, respectively. The red dots inscribed in the red-shaded domain correspond to the simulated conditions where $|\Delta G_{HB}| > |\Delta G_{field}|$ and water does not align to the field. 
At weakly interacting surfaces, in the range of interaction strengths typical of, e.g., commonly studied carbon electrodes, the red-shaded domain vanishes and water alignment simplifies to the traditional one-dimensional problem as a function of the applied voltage. By contrast, the widening of the red-shaded domain with increasing surface--water interactions (increasing surface hydrophilicity) shows the importance of understanding the uncovered $\Delta G_{HB}$ driving force for strongly interacting metal electrodes.

To this end, we propose the following conditioned function to generally describe water alignment -- as probed by $\chi^{(2)}$  intensities in SFG/SHG experiments  -- in the whole phase diagram:
\begin{equation}\label{eq:conditioned}
 \chi^{(2)}(V) - \chi^{(2)}(PZC) \simeq
\begin{cases}
0  & \text{if } V < V^*(\epsilon)\\
\beta^{eff}(\sum_{i=1}^{N{_{ADL}}} cos\theta_i(V) - \sum_{j=1}^{N{_{ADL}}}  cos\theta_j(PZC)) & \text{if }  V \geq V^*(\epsilon)
\end{cases}
\end{equation}
%
%


%


where $V^*(\epsilon)$ is the applied voltage at which $\Delta G_{field} = - \Delta G_{HB}$, which is a function of surface-water interactions ($\epsilon$), according to Eq.\ref{eq:V*}. For $V < V^*(\epsilon)$, $\Delta G_{HB}$ dominates and water does not align, while the surface field is larger otherwise, enhancing the spectroscopic response (according to Eq. \ref{eq:SFG-ADL-diff}).

With Eq. \ref{eq:conditioned}, we can finally understand the results of the SHG experiments at a Ni electrode from ref.\citenum{speelman2025quantifying}. We show this in Fig.\ref{surface_effect}B, where we use Eq. \ref{eq:conditioned} to quantitatively model the water alignment as measured by the SHG response, without simulations of the Ni/water interface. We start by considering that Ni interacts more strongly then Au with water,\cite{heinz2008accurate, eggert2023cavity,Michaelidis_ChemRev2016} which is expected to result in a larger $\Delta G_{HB}$, and hence $V^*$. Moreover, this leads to surface oxidation, which likely changes the surface ordering of the water network compared to the horizontal ordering at the bare metal electrode. This offers a nice opportunity to test the generality of our physical model for water alignment. It is important to note that $\Delta V$ in the Ni experiment is referenced to OCP, instead of PZC. We take this into account by fitting $V^*$  to the experiments. At  $V > V^*$, the spectroscopic response is expected to be dominated by the applied voltage contribution $\Delta G_{field}$, according to our model. Therefore, we take the $\sum_{i=1}^{N{_{ADL}}} cos\theta_i(V)$ term in Eq.\ref{eq:conditioned} from our Au/water simulations (shifted in V by the difference in V$^*$ between Au and Ni). The obtained theoretical voltage dependent spectroscopic response matches well the experiments. In particular, the voltage range where water is not aligned and the spectroscopic response is zero, from -0.6 V to 0.2 V, is larger (by a factor of $\sim$1.5) compared to Au (Fig.\ref{effects}A), corresponding to a larger $\Delta G_{HB}$, in nice agreement with the expectation from our physical model.

\section{Conclusion}\label{sec13}
In summary, with the help of plasmon-enhanced SFG spectroscopy and constant-potential MD simulations, we proposed a physical model that reconciles the traditional picture of water alignment at electrodes with the recent results on metal electrodes from SHG and SFG non-linear optical techniques.\cite{speelman2025quantifying, w1} Contrary to textbook descriptions, water does not align immediately upon applying voltage at these metal electrodes. Instead, alignment only begins beyond a surface-dependent threshold voltage, due to the ordering of interfacial water network templated by the metal. We identified two competing free energy contributions governing water alignment: a penalty to rearrange the H-bond network templated by the surface ($\Delta G_{HB}$), and the canonical voltage-dependent driving force ($\Delta G_{field}$) to align water. Through a theoretical experiment in which we varied the surface–water interaction strength, we demonstrated that stronger surface–water interactions increase $\Delta G_{HB}$, thus requiring higher voltages to overcome the H-bond resistance and induce alignment. For weakly interacting surfaces, $\Delta G_{HB}$ vanishes and we recover the canonical behavior previously observed at carbon electrodes. Our physical model can be represented by a phase diagram that describes water alignment in the whole range from weakly-interacting carbon electrodes, for which the textbook picture applies, to metal electrodes. For metals, $\Delta G_{HB}$ and $\Delta G_{field}$ are of comparable magnitude in a wide voltage range, until the chemistry starts both at the positive and at the negative electrode. This suggests that $\Delta G_{HB}$ may play a major role in controlling water alignment and the reorganization of the interfacial H-bond network in electrochemical reactions, such as HER and OER.

\section{Methods}\label{sec11}
\textbf{Computational methods.} The constant potential classical MD code MetalWalls~\cite{marin2020metalwalls} was used for all simulations, with box dimensions 3.66$\times$3.66$\times$7.00 nm, containing 2,381 SPC/E~\cite{berendsen1987missing} water molecules confined between two identical planar Au(100) electrodes, each composed of five atomic layers. Periodic boundary conditions were applied in the $x$ and $y$ directions. The Lennard-Jones parameters for Au were adopted from Heinz et al.~\cite{heinz2008accurate} and combined with Lorentz-Berthelot mixing rules. The system was first equilibrated in the NVT ensemble at $T = 298$ K for 500 ps with a timestep of 1 fs, followed by 500 ps where the electrodes acted as pistons to apply a constant pressure of 1 atm until the system's density stabilized. Subsequent production NVT runs were performed with a 2 fs timestep for at least 30 ns. We employed a Nosé–Hoover thermostat~\cite{martyna1992nose} chain with a 1 ps time constant. Since the SFG experiments were conducted at pH 12, corresponding to 10$^{-2}$ M ionic strength, on a multicrystalline electrode, we confirmed the generality of our findings by additional MD simulations (with the same protocol) with varying surface topology (111 instead of 100) and ionic strength (by adding 0.1 M NaCl, with force field parameters for NaCl from ref.\cite{Kann2014}).  For H-bond analyses, we employed a distance$+$angle criterion with 3.5 \AA~ and $\pm 30^o$ (we confirmed that similar trends are obtained by slightly varying the cutoff). 
\\


\section{Acknowledgements}

This work is funded by the European Research Council (ERC, ELECTROPHOBIC, Grant Agreement No. 101077129) . This work was performed using HPC resources from GENCI, allocation No. A0170802309. We thank Dr. Wanlin Chen for The Interesting And Motivating Opinions shared. 

\noindent

\bigskip
\bibliography{references}

\end{document}